\documentclass[aps,pra,preprint,groupedaddress,showpacs,superscriptaddress]{revtex4-1}
\usepackage{amssymb, amsmath, bm, braket, graphicx, color}
\usepackage[normalem]{ulem}

\begin{document}

\title{Light-dressed states under intense optical near fields}

\author{Takashi Takeuchi}
\email{take@ccs.tsukuba.ac.jp}
\affiliation{Center for Computational Sciences, University of Tsukuba, Tsukuba 305-8577, Japan}
\author{Tokuei Sako}
\email{sako.tokuei@nihon-u.ac.jp}
\affiliation{Laboratory of Physics, College of Science and Technology, Nihon University, 7-24-1 Narashinodai, Funabashi, Chiba 274-8501, Japan}
\author{Katsuyuki Nobusada}
\email{nobusada@ims.ac.jp}
\affiliation{Department of Theoretical and Computational Molecular Science, Institute for Molecular Science, Myodaiji, Okazaki 444-8585, Japan}

\date{\today}

%\TS{The text added/changed by TS is colored red.}
%\TSS{The comment added by TS is colored blue.}

\begin{abstract}
We studied the excitation dynamics of a finite quantum system with an intense optical near field (ONF) from a perspective of light-dressed states. A simple model consisting of a single electron and a nano-sized short dipole source were employed. By calculating the time-dependent wave function subjected to the ONF, we demonstrated that the optical responses involved not only the first- and third-order but also the second-, fourth-order harmonic generations that were not obtained from conventional spatially homogeneous laser fields. In order to elucidate the origins of the exotic high-order harmonic generations, the dressed states altered by the ONF were explored. The result showed that the spatial distribution of the dressed states were significantly influenced by the ONF, making forbidden optical transitions to be allowed by parity breaking. This was caused by the spatial inhomogeneousness of the ONF, specifically its asymmetry, that was inherited to the dressed states. 
\end{abstract}

\maketitle

%%%%%%%%%%%%%%%%%%%%%%%%%%%%%%%%%%%%%%%%%%%%%%%%%%%%%%%%%%%%%%%%%%%
\section*{I. INTRODUCTION}
%%%%%%%%%%%%%%%%%%%%%%%%%%%%%%%%%%%%%%%%%%%%%%%%%%%%%%%%%%%%%%%%%%%

A finite quantum system, such as an atom or a molecule, exposed to an intense laser field more than $10^{12}$  W/cm$^{2}$ reforms itself as new quantum states, called light-dressed states or simply dressed states \cite{dre_rev_01, dre_rev_02, dre_rev_03}. In dressed states electrons in the matter and the incident laser light are strongly coupled, which allows them to have new quai-eigen energy levels and probability-density distributions distinct possibly from those of the original uncoupled states through repetitive photon absorptions and emissions. 

Theoretical studies on dressed states initiated in the late 1970s \cite{dre_theo_01}, where resonance fluorescence and absorption spectra of a multi-level atom was investigated. The approach determines the quasi-eigen energy levels of the system of coupled light and electrons as eigenvalues of the Floquet Hamiltonian \cite{dre_floq_01}. Since then, the dressed-state approach has been applied in analyzing a wide variety of novel physical phenomena in laser physics and chemistry or in quantum optics, such as dynamical Stark effect \cite{dre_stark_01, dre_stark_02, dre_stark_03}, electromagnetically induced transparency \cite{dre_eit_01, dre_eit_02, dre_eit_03}, cavity QED \cite{dre_ccqed_01, dre_ccqed_02, dre_ccqed_03}, and so on. Furthermore, recent advances in strong- and/or ultrashort-pulsed laser light technology have opened new sophisticated ways of manipulating atoms and molecules relying on dressed states, demonstrated such as in softening or hardening of chemical bonds in photochemical reactions \cite{dre_chembond_01, dre_chembond_02, dre_chembond_03, dre_chembond_04} or in laser-assisted elastic electron scattering \cite{dre_obs_01}. 

In most of these pioneering previous studies dealing with dressed states spatial homogeneousness in the external laser field has been assumed. This assumption together with a commonly used condition in atomic and molecular physics, asserting that the size of the target electron system is by far much smaller than the typical wavelength of the laser lights, leads us to use the dipole approximation safely. Meanwhile, spatially {\it inhomogeneous} electromagnetic fields have attracted increasing attention in the last few decades. A most notable example of such an inhomogeneous field playing an important physical role may be an optical near-field (ONF) \cite{onf_rev_01, onf_rev_02, onf_rev_03} that is a localized electromagnetic field generated by the induced surface charge of the nanostructured material interacting with the incident laser light. Since the spatial locality of the ONF depends on the size of the nano material, it can break the diffraction limit of light and can reach as small as even the atomic scale \cite{np_tddft_01}. This locality has been widely utilized in the area of nanophotonics, such as in highly integrated optical signal-processing circuits \cite{onf_circuit_01, onf_circuit_02, onf_circuit_03} or in near-field scanning optical microscopy \cite{onf_scopy_01, onf_scopy_02, onf_scopy_03}. 

From a theoretical point of view, since the ONF is spatially inhomogeneous due to its high locality, we could no longer rely on the dipole approximation. This implies that the optical responses of the target electron system interacting with the ONF can be significantly different from predictions based on conventional theoretical models assuming homogeneous electromagnetic fields. Indeed, recent studies, both experimentally and theoretically, have shown that the inhomogeneousness of the ONF inherently induces electric-quadrupole transitions \cite{onf_qua_01, onf_qua_02, onf_qua_03, onf_qua_04} and second harmonic generation (SHG) \cite{onf_shg_01, onf_shg_02, onf_shg_03} for symmetric materials, the latter of which is forbidden under standard homogeneous electromagnetic fields. In weak-field cases a recent theoretical study based on a perturbation theory has demonstrated that the inhomogeneousness in the ONF, specifically its electric field gradient, is responsible for the aforementioned exotic optical responses \cite{onf_shg_02}. Considering the current situation of growing interests in strong laser fields \cite{onf_strong_01, onf_strong_02, onf_strong_03, onf_strong_04, onf_strong_05, onf_strong_06}, a general theoretical model for the ONF that is applicable to both inhomogeneous and strong fields beyond the perturbative regime needs to be developed.

In the present study we have investigated the excitation dynamics of a finite electron system with an intense ONF from a perspective of the dressed states. This paper is organized as follows. The next section describes our theoretical model and computational details. The first subsection starts with the definition of the studied system. The system shows to be modeled by a single electron confined in a quasi-one-dimensional potential well, that mimics a quantum wire \cite{qd_01, qd_02}, while the ONF is described by a short dipole source. To obtain the temporal evolution of the system we have chosen to solve the time-dependent Schr\"{o}dinger equation directly relying on the finite-difference time-domain (FDTD) scheme \cite{fdtd_quan_01}. The next subsection in Sec.~II describes a brief introduction of the computational aspects of dressed states. We show here the relation between the time-dependent solution of the Schr\"{o}dinger equation and the time-independent solution of the Floquet matrix. The computational results and their interpretation are given in Sec.~III. The optical responses to the ONF are shown here as the dynamic induced dipole moment obtained from the time-dependent wave function. Fourier transformation of the induced dipole moment for the results obtained by exciting the system with a spatially asymmetric ONF shows several frequency components distinct from those corresponding to the dipole-allowed transitions. To rationalize the appearance of these exotic spectral lines we construct a theoretical model of dressed states for a system subjected to an inhomogeneous ONF. The quasi-eigenenergies obtained by solving the eigenvalue equation for the Floquet Hamiltonian \cite{dre_floq_01} are shown to assign each spectral line including these exotic ones. The probability distributions of the eigenstates of the Floquet Hamiltonian for the asymmetric ONF also show appreciable asymmetry, confirming that this broken parity makes a number of forbidden transitions to be active and thus allows the new optical responses of the SHG, fourth harmonic generation (FHG), and difference-frequency generation (DFG) spectra. The main results of the present study is summarized in Sec.~IV.

%%%%%%%%%%%%%%%%%%%%%%%%%%%%%%%%%%%%%%%%%%%%%%%%%%%%%%%%%%%%%%%%%%%
\section*{II. THEORETICAL MODEL AND COMPUTATIONAL DETAILS}
%%%%%%%%%%%%%%%%%%%%%%%%%%%%%%%%%%%%%%%%%%%%%%%%%%%%%%%%%%%%%%%%%%%
\subsection*{A. Time-dependent Schr\"odinger equation}%********************************************************************************%
\begin{figure}[htbp]
\centering
\includegraphics[width=16cm, trim=0 200 0 100]{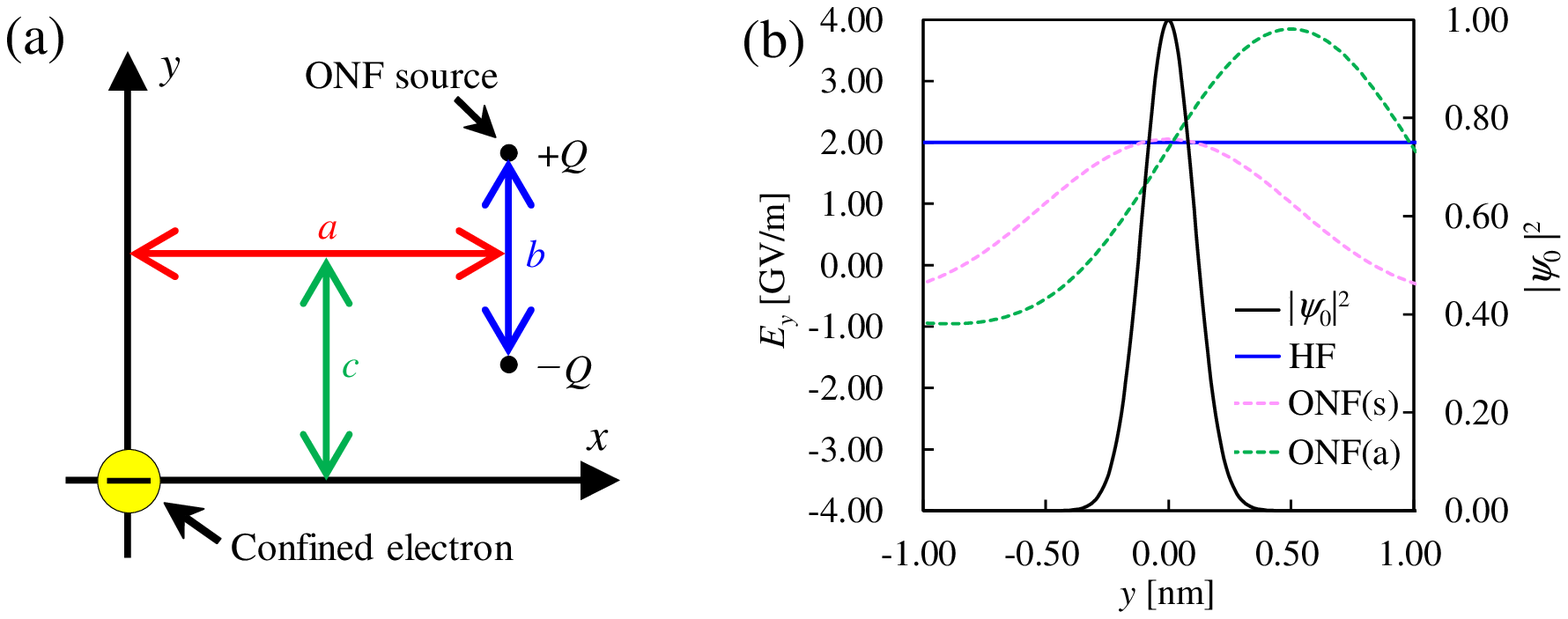}
\caption{(Color online) (a) A schematic illustration of the studied system consisting of a confined electron and an ONF source with the definition of the distance parameters $a$, $b$ and $c$ (in nm) specifying relative positions of the electron and the ONF source. The single electron is confined in a quasi-one-dimensional potential well extending along the $y$-axis (see Eq.~\eqref{vsta}). (b) The probability density distribution of the ground electronic state $\psi_0$ and three different electric fields along the $y$-axis, displayed, respectively, by the curved solid black, flat solid blue (black), broken pink (light gray), and broken green (dark gray) lines. The ordinate on the left-hand side represents the probability density of the electron (in nm$^{-1}$) while that on the right-hand side represents the electric-field strength (in GV/m). HF, ONF(s), and ONF(a) represent, respectively, a homogeneous field corresponding to a conventional laser light, a symmetric ONF, and an asymmetric ONF (See text). The parameters for ONF(s) and ONF(a) are $(a, b, c) = (1, 1, 0)$ and $(1, 1, 0.5)$, respectively.}
\label{fig:model}
\end{figure}

We consider a model system of an electron interacting with an ONF source, which is schematically illustrated in Fig. \ref{fig:model} (a). The single electron is assumed to be confined in a quasi-one-dimensional potential well extending along the {\it y} axis. The time-dependent Schr\"{o}dinger equation subject to an external electromagnetic field used in the present study is described by
\begin{align}
i \hbar \frac{\partial \psi(y,t)}{\partial t} = \left[
-\frac{\hbar^{2}}{2m}\frac{\partial^2}{\partial y^2}  +V_{sta}(y)+V_{ext}(y,t)
\right]\psi(y,t),
\label{tdse}
\end{align}
where $ V_{sta} $ and $ V_{ext} $ are the electrostatic confinement potential and the external electromagnetic potential, respectively. 
We employ the following soft coulomb potential for $ V_{sta} $,
\begin{align}
V_{sta}(y) = \alpha\left( \frac{1}{\beta}-\frac{1}{\sqrt{y^{2}+\beta^{2}}} \right),
\label{vsta}
\end{align}
where $ \alpha $ and $ \beta $ are chosen to be 32.5 eV $ \cdot $ nm and 1.75/3 nm, respectively. These parameters allow us to set the excitation energy of the system into the visible light range. The ground state $ \psi_{0} $ of this potential is chosen as the initial state for the time-dependent analysis. The probability density distribution of $ \psi_{0} $ is displayed in Fig.~\ref{fig:model}(b). 

For the external electromagnetic potential $ V_{ext} $, we have the following seperable form for its spatial and time dependences as
\begin{align}
V_{ext}(y,t) = V_{s}(y)V_{t}(t).
\label{vext}
\end{align}
For the temporal dependence we employ a simple sinusoidal form for $ V_{t}(t) $ with a step function $ u(t) $ described by
\begin{align}
V_{t}(t) = \sin(\omega_{i}t)u(t),
\label{td_vt}
\end{align}
where the angular frequency $\omega_{i}$ of the incident laser light is chosen as $\hbar\omega_i =$ 3.62 eV so that it corresponds to 1.1 times the excitation energy $\hbar\omega_{01}$ between the ground $ \psi_{0} $ state and the first-excited $ \psi_{1} $ states. This factor of 1.1 is introduced to avoid strong resonance between these two original eigenstates, that allows us to identify each dressed state easily from the computational results. 

For the spatial dependence in $V_{ext}$ of Eq.~(\ref{vext}) we have examined the following three types of excitation schemes by changing the function form of $ V_{s} $: First, we have considered a conventional laser field as a reference to homogeneous-field cases, which is described in the length gauge \cite{length_gauge_01} as
\begin{align}
V_{s}^{hf}(y) =qE_{0}y,
\label{vhf}
\end{align}
with $q$ and $E_0$ denoting the charge of the particle and the amplitude of the electric field, respectively. The gradient of this potential along the $ y $ axis provides a homogeneous electric field. The amplitude $ E_{0} $ has been set to 2 GV/m that corresponds to about $10^{12}$ W/cm$^{2}$ in the laser intensity. Second, we have introduced a short dipole source to produce an ONF as represented in Fig.~\ref{fig:model}(a). This dipole source has the length $ b $ and is situated away from the confined electron whose location is specified by the distance parameters $ a $ and $ c $. When the dipole length $ b $ is sufficiently shorter than the wavelength of the incident laser light (that is 342 nm for $ \omega_{i} $), the electromagnetic field reaches a state of quasi-static \cite{quasi_sta_01}. The scalar potential is then dominant and the function $ V_{s} $ is described with a charge $ Q $ induced at the ends of the dipole as
\begin{align}
V_{s}^{onf}(y) =\frac{Q}{4\pi\epsilon_{0}}\left[  
\frac{1}{\sqrt{a^{2}+ \left\{ y- \left( c+\frac{b}{2} \right) \right\}^{2} }} 
-\frac{1}{\sqrt{a^{2}+ \left\{ y- \left( c-\frac{b}{2} \right) \right\}^{2} }} 
\right],
\label{vonf}
\end{align}
where we assume the Coulomb gauge for the ONF excitation. To make a fair comparison with the result by the conventional excitation scheme with $V_{s}^{hf}$ the charge $ Q $ has been determined by the following equation so as to keep the same electromagnetic energy:
\begin{align}
\int |\psi_{0}(y)|^{2} \left| \frac{\partial V_{s}^{onf}(y)}{\partial y} \right|^{2}dv
= \int |\psi_{0}(y)|^{2} \left| \frac{\partial V_{s}^{hf}(y)}{\partial y} \right|^{2}dv.
\label{qchoose}
\end{align}
By introducing this condition the intensity of the ONF maintains $10^{12}$  W/cm$^{2}$. In this ONF excitation we can introduce a symmetric and asymmetric field to the confined electron by choosing zero or nonzero value of the $c$ parameter as seen from Eq.~(\ref{vonf}) or Fig.~\ref{fig:model}(a). In Fig.~\ref{fig:model}(b) we have plotted the electric field along the $y$-axis for the conventional field with $ V_{s}^{hf} $ (denoted hereafter by HF), the symmetric and asymmetric ONFs, denoted respectively by ONF(s) and ONF(a), where we have employed $ a $=$ b $=1 nm and $ c $=0 nm for ONF(s) while $ a $=$ b $=1 nm and $ c $=0.5 nm for ONF(a). As shown in this figure HF is a constant and thus a spatially homogeneous field as its name indicates, while both ONF(s) and ONF(a) have nonzero $y$ dependences and are thus inhomogeneous-symmetric and inhomogeneous-asymmetric fields, respectively. 

Under the above three excitation schemes, namely, HF, ONF(s), and ONF(a), we have calculated the time evolution of the electronic wave function by using the FDTD method. The FDTD method is originally developed for solving Maxwell's equations and requires us to spatiotemporally discretize the electric and magnetic fields \cite{fdtd_01, fdtd_02}. For the Schr\"{o}dinger equation, by dividing the wave function into the real and imaginary parts, we can obtain recursive equations and stably update the electron wavepacket discretized in the spatial-grids. The detailed numerical implementation has been described in our previous studies \cite{fdtd_quan_01, fdtd_quan_02}.

\subsection*{B. Dressed state under optical near field}%**********************************************************************************%
According to Floquet's theorem \cite{dre_floq_01}, when $ V_{ext} $ has perfect time-periodicity with the period $T_i = 2\pi/\omega_i$, such as, 
\begin{align}
V_{ext}(y,t) = V_{s}(y)\left\{ \frac{\exp(i\omega_{i}t)+\exp(-i\omega_{i}t)}{2} \right\},
\label{drevext}
\end{align}
the time-dependent wave function $ \Psi $ can be factorized into a product of a time-dependent phase factor with a quasi-eigenenergy $\epsilon$ and a time-periodic part $\Pi$ as
\begin{align}
\Psi(y,t)=\exp\left( -i\frac{\epsilon}{\hbar}t \right)\Pi(y,t),
\label{dresolu}
\end{align}
where $\Pi$ satisfies the condition, $\Pi(y,t+T_i) = \Pi(y,t)$. Then, $\Pi$ can be represented by a Fourier series as
\begin{align}
\Pi(y,t)=\sum_{n=-N}^{N}\zeta_{n}(y)\exp(i\omega_{i}nt).
\label{dresolu_time}
\end{align}
In Eq.~(\ref{td_vt}) the time-dependence of $V_{ext}$ has been defined by a sine function multiplied by a step function. We have however assumed in Eq.~(\ref{drevext}) a cosine function. In the time-dependent calculation the choice of the functions, sine or cosine, would certainly affect the result owing to their difference in the initial value at $t = 0$. In the case of the following Floquet analysis, on the other hand, the result wouldn't change by this choice since it is a stationary time-independent calculation. We have therefore chosen a simpler cosine function rather than sine, that can give a real-symmetric representation of the Floquet matrix that would otherwise become Hermitian complex.

Mathematically the summation in this Fourier expansion of Eq.~(\ref{dresolu_time}) extends from $-\infty$ to $\infty$, but we have limited $n$ between $-N$ to $N$ so as to perform practical calculation. The index $n$ can be interpreted as the number of photons coupled to the electron system. For the spatial part for each $n$, namely $\zeta_n$, it is expanded by the eigen functions $ \psi_{k} $ bound in the electrostatic potential $ V_{sta} $ as
\begin{align}
\zeta_{n}(y)=\sum_{k=0}^{K}\zeta_{k}^{n}\psi_{k}(y),
\label{dresolu_spa}
\end{align}
where the integer $K$ denotes the index for the highest eigenstate.

\begin{figure}[htbp]
\centering
\includegraphics[width=14cm, trim=0 200 0 100]{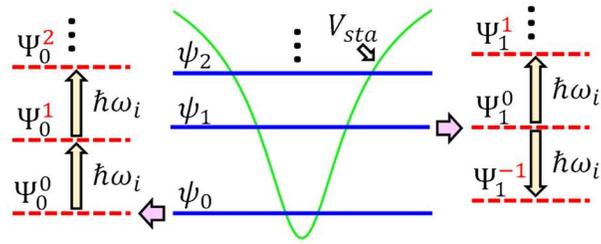}
\caption{(Color online) A schematic illustration of the formation of the zeroth-order dressed states $\Psi_{k}^{n}$ from the original electronic states $\psi_k$ (See Eq.~(\ref{dresolu_spa})). The solid curve and the solid bars attached to it displayed at the center represent, respectively, the electrostatic potential $ V_{sta} $ and its eigenenergy levels corresponding $ \psi_{k} $. On the left- and right-hand sides of the figure the energy levels of the dressed states $ \Psi_{k}^{n}$ with the photon number $n$ ($n = \cdots, -1, 0, 1, \cdots$) (the broken bars) originating from the ground $\psi_{0}$ state and the first-excited $\psi_1$ state are displayed.}
\label{fig:schematic}
\end{figure}

By substituting Eqs. \eqref{dresolu}~-~\eqref{dresolu_spa} into Eq. \eqref{tdse}, multiplying $\psi_l$ from the left and integrating it with respect to $y$, we obtain the following equation:
\begin{align}
\sum_{k=0}^{K}\zeta_{k}^{n+1}V_{l,k}+\zeta_{l}^{n}(e_{l}+\hbar\omega_{i}n)+\sum_{k=0}^{K}\zeta_{k}^{n-1}V_{l,k}=\epsilon\zeta_{l}^{n},
\label{dreeq1}
\end{align}
where $V_{l,k}$ and $e_l$ are defined, respectively, by
\begin{align}
V_{l,k}=\frac{1}{2}\Braket{\psi_{l}|V_{s}(y)|\psi_{k}},
\label{dreeq2}
\end{align}
and
\begin{align}
e_l=\Braket{\psi_{l}|\left[-\frac{\hbar^{2}}{2m}\frac{\partial^{2}}{\partial y^{2}}  +V_{sta}(y)\right]|\psi_{l}}.
\label{dreeq3}
\end{align}
Eq. \eqref{dreeq1} can be written in the following form of a matrix eigenvalue problem as
\begin{align}
\left[
        \begin{array}{ccccccc}
         \bar{H}_{-N} & \bar{V} & 0 & 0 & 0 & 0 & 0 \\
         & \ddots & & & & & \\
         0 & \bar{V} & \bar{H}_{n-1} & \bar{V} & 0 & 0 & 0 \\
         0 & 0 & \bar{V} & \bar{H}_{n} & \bar{V} & 0 & 0  \\
         0 & 0& 0 & \bar{V} & \bar{H}_{n+1} & \bar{V} & 0 \\
         & & & & & \ddots & \\
         0 & 0 & 0 & 0 & 0 & \bar{V} & \bar{H}_{+N} \\
         \end{array}
\right]
\left[
         \begin{array}{c}
         \bar{\zeta}_{-N} \\
         \vdots \\
         \bar{\zeta}_{n-1} \\
         \bar{\zeta}_{n} \\
         \bar{\zeta}_{n+1} \\
         \vdots \\
         \bar{\zeta}_{+N} \\
         \end{array}
\right]=\epsilon\left[
         \begin{array}{c}
         \bar{\zeta}_{-N} \\
         \vdots \\
         \bar{\zeta}_{n-1} \\
         \bar{\zeta}_{n} \\
         \bar{\zeta}_{n+1} \\
         \vdots \\
         \bar{\zeta}_{+N} \\
         \end{array}
        \right],
\label{dreeq4}
\end{align}
where $\bar{V}$ and $\bar{H}_n$ are $(K+1){\times}(K+1)$ square matrices with $\bar{\zeta}$ denoting a vector of length $K+1$ defined, respectively, by
\begin{align}
\bar{V}=\left[
        \begin{array}{ccc}
         V_{0,0} & \cdots & V_{0,K}  \\
         \vdots & \ddots & \vdots \\
         V_{K,0} & \cdots & V_{K,K} \\
         \end{array}
         \right],
\label{dreeq5}
\end{align}
\begin{align}
\bar{H}_{n}=\left[
        \begin{array}{ccc}
         e_0+\hbar\omega_{i}n & 0 & 0  \\
         0 & \ddots & 0 \\
         0 & 0 & e_K+\hbar\omega_{i}n \\
         \end{array}
         \right],
\label{dreeq6}
\end{align}
and
\begin{align}
\bar{\zeta}_{n}=\left[
         \begin{array}{c}
         \zeta_{0}^{n} \\
         \vdots \\
         \zeta_{K}^{n} \\
         \end{array}
         \right].
\label{dreeq7}
\end{align}
By solving the eigenvalue equation~\eqref{dreeq4} the dressed-state wave function $ \Psi $ and its quasi-eigenenergy $\epsilon$ are obtained \cite{dre_floq_01,dre_floq_02}. Dressed states in their zeroth-order are a direct product of an electronic state $\psi_k$ and a photon number state with its energy $\hbar{\omega_i}n$. Fig.~\ref{fig:schematic} shows schematically how these zeroth-order dressed states are formed from the original electronic states. In the middle of this figure the energy levels of the original electronic states $\psi_k$ bound to the confinement potential $V_{sta}$ are displayed while on its both sides the energy levels of the zeroth-order dressed states $\Psi_{k}^{n}$ generated from $\psi_k$ are plotted for the ground and the first-excited states, $ \psi_{0} $ and $ \psi_{1} $, respectively. Since nearby energy levels $\Psi_{k}^{n}$ and $\Psi_{k}^{n+1}$ have the energy difference $ \hbar\omega_{i} $, $\Psi_{k}^{n}$ is interpreted as an $n$-photon coupled state with respect to $\Psi_{k}^{0}$. These zeroth-order states are coupled through the interaction term $V_{s}$. When the light intensity is not too high, the dressed-state wave functions can keep their original character of uncoupled states, allowing us to identify the original electronic state $\psi_k$ and the number of coupled photons. Since in our present case of $10^{12}$ W/cm$^{2}$ each dressed-state wave function is dominated by its zeroth-order state $\Psi_{k}^{n}$, we have used the same notation $\Psi_{k}^{n}$ to indicate the dressed eigenstates.

The Floquet matrix of Eq.~(\ref{dreeq4}) is of size $(2N+1)(K+1)$, which gives a total of $(2N+1)(K+1)$ eigenvalues and eigenvectors. However, since the Floquet matrix should be of an infinite dimension in a rigorous sense, eigenvalues of its truncated matrix like that in Eq.~(\ref{dreeq4}) might suffer from incomplete convergence, particularly those eigenvalues located close to the top or bottom on the energy axis. On the other hand, eigenvalues located close to the center, that have a small number of coupled photons, are in general fast in convergence for increasing $N$ and we have focused on them in our present analysis.

%%%%%%%%%%%%%%%%%%%%%%%%%%%%%%%%%%%%%%%%%%%%%%%%%%%%%%%%%%%%%%%%%%%
\section*{III. RESULTS AND DISCUSSION}
%%%%%%%%%%%%%%%%%%%%%%%%%%%%%%%%%%%%%%%%%%%%%%%%%%%%%%%%%%%%%%%%%%%

\begin{figure}[htbp]
\centering
\includegraphics[width=16cm, trim=0 200 0 100]{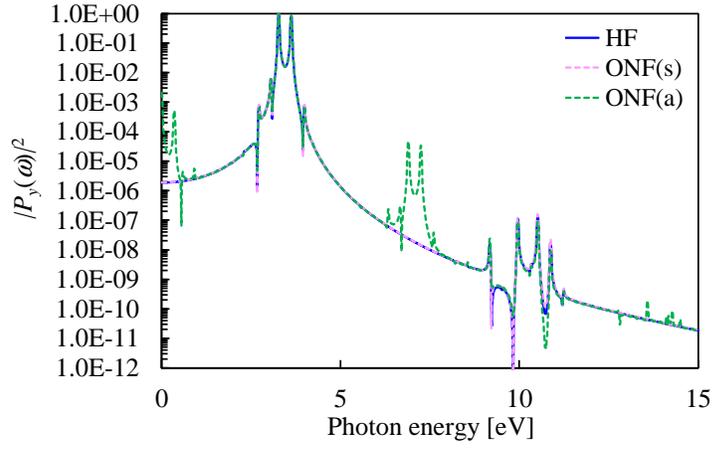}
\caption{(Color online) Normalized power spectra of the induced dipole moment $ |P_{y}(\omega)|^{2} $ (See Eq.~(\ref{ind_p})) calculated from the time-dependent electron wave packets. Results obtained by HF, ONF(s), and ONF(a) (see text for their definition) are shown by the solid blue (black), broken pink (light gray), and broken green (dark gray) lines, respectively.}\label{fig:power_spe}
\end{figure}

First, we have calculated the time-evolution of the electron wave packet subjected to an external electromagnetic field in three different excitation schemes, namely, HF, ONF(s), and ONF(a), respectively, as shown in Fig. \ref{fig:model} (b). The spacing of the space and time grids has been chosen as ${\Delta}x = {\Delta}y =$ 17.5 pm and ${\Delta}t =$ 0.24 as. From the time-dependent wave packets we have calculated power spectra of the induced dipole moment $P_y$ defined by
\begin{align}
P_{y}(\omega)=F\left[ f(t)\int qy (|\psi(y,t)|^{2}-|\psi_{0}(y)|^{2}) dv \right],
\label{ind_p}
\end{align}
where $F$ and $ f(t) $ denote, respectively, the time-frequency Fourier transformation and a window function of the form,
\begin{align}
f(t)=1-3\left( \frac{t}{T_{max}} \right)^{2}+2\left( \frac{t}{T_{max}} \right)^{3},
\label{wf}
\end{align}
with $ T_{max} $ being set to 115 fs. This window function is introduced to avoid spurious peaks in the transformation. The power spectrum of Eq.~(\ref{ind_p}) is normalized in the sense that the intensity of its maximum peak is equal to 1. The results have been displayed in Fig.~\ref{fig:power_spe}  by the solid blue (black), broken pink (light gray), and broken green (black) lines obtained, respectively, by the HF, ONF(s), and ONF(a) excitation schemes. As shown in this figure the results for HF and ONF(s) look similar: They both have characteristic peaks at around 3.6 eV and 11 eV, which correspond, respectively, to the first- and third-harmonic generations of $ \omega_{i} $. These harmonic generations follow the selection rules based on the dipole approximation. In other words, the spatial inhomogeneousness of ONF(s) has hardly contributed to the optical response. In contrast, the result for ONF(a), plotted by the broken green (dark gray) line, shows additional peaks at around 0.3, 7.2, and 14 eV. Those emerging peaks coincide, respectively, with the frequencies by DFG between $ \omega_{i} $ and $ \omega_{01} $ (the energy gap between $ \psi_{0} $ and $ \psi_{1} $), i.e. 0.1$ \omega_{01} $, SHG, and by FHG. This result indicates that the inhomogeneousness of ONF(a), in particular its asymmetry, breaks down the dipole approximation, causing the exotic optical responses.

\begin{figure}[htbp]
\centering
\includegraphics[width=16cm, trim=0 200 0 100]{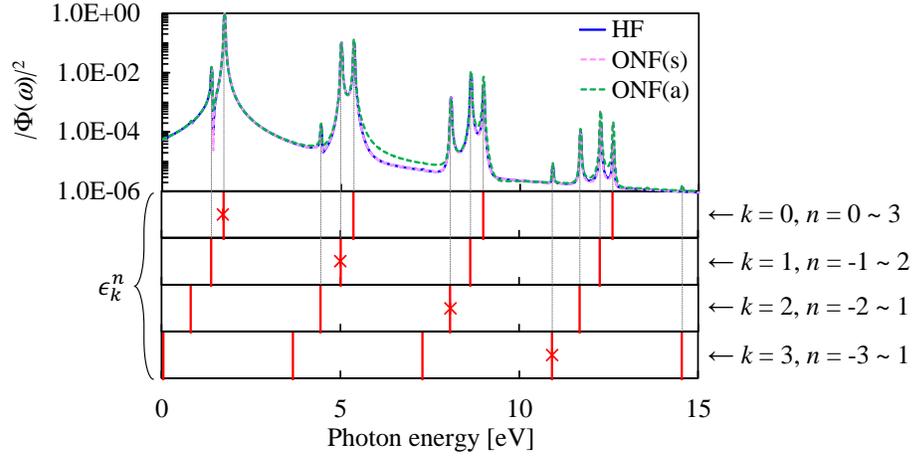}
\caption{(Color online) Normalized power spectra of the time-dependent wave function $ |\Phi(\omega)|^{2} $ (the upper panel, see Eq. \eqref{psi_spe}) and dressed energy levels $ \epsilon_{k}^{n} $ represented by the red (black) solid vertical lines (the lower panels). The cross marks indicate the zero-photon-coupled reference state $\epsilon_{k}^{0}$ for each $ k $. The spectra obtained by HF, ONF(s), and ONF(a) (see text) are shown by the solid blue (black), broken pink (light gray), and broken green (dark gray) lines, respectively.
}\label{fig:power_spe_phi}
\end{figure}

In order to elucidate the origin of the DFG, SHG, and FHG by ONF(a) we have calculated and analyzed the dressed states for the studied system. The upper panel of Fig.~\ref{fig:power_spe_phi} represents normalized power spectra $\Phi(\omega)$ of the time-dependent wave function $\psi(t)$ defined by
\begin{align}
\Phi(\omega)=\int F[f(t)\psi(y,t)] dv.
\label{psi_spe}
\end{align}
In this panel, as has been done in Fig. \ref{fig:power_spe}, the solid blue (black), broken pink (light gray), and broken green (dark gray) lines represent, respectively, the results obtained by HF, ONF(s), and ONF(a). In contrast to the results of the induced dipole moment shown in Fig. \ref{fig:power_spe}, the power spectra of the time-dependent wave functions by the three distinct excitation schemes yield almost the same spectra. Each peak in the spectra has its origin in the dressed\-state eigenenergies $ \epsilon_{k}^{n} $, represented by the red (black) solid vertical lines at the lower four panels of the figure. Since $ \epsilon_{k}^{n} $ is basically determined by the sum or difference between the original eigenenergies of $ V_{sta} $ and the photon energies ${\hbar\omega}n$, which is irrespective of the way of excitations, we have plotted in the lower panels only the results obtained by HF. The eigenenergies $\epsilon_{k}^{n}$ in this figure have been obtained by diagonalizing a truncated Floquet matrix in Eq.~(\ref{dreeq4}) constructed with the condition $(K,N) = (20,10)$. We have selected those eigenstates originating from $ k = 0\sim3 $ by analyzing the eigenvectors. The cross marks in the figure indicate the eigenenergy $ \epsilon_{k}^{0} $ of the `zero-photon coupled state' for each $ k $. Consequently, energy levels on the right-hand side from the marks indicate states that have absorbed photon(s) with a positive $n$ number with respect to the reference state $\Psi_{k}^{0}$, while those on the left-hand side are states that have emitted photon(s) with a negative $n$ value. The correspondence between each peak in the spectra and the dressed-state eigenenergy $\epsilon_{k}^{n}$ has been made and indicated by the gray broken vertical lines  drawn from the upper to lower panels. All spectral lines in $ \Phi $ have shown to be able to find a corresponding counterpart in the eigenenergies $ \epsilon_{k}^{n} $. In particular, $ \epsilon_{0}^{0\sim2} $ and $ \epsilon_{1}^{-1\sim1} $ have shown to dominate the spectra.

\begin{figure}[htbp]
\centering
\includegraphics[width=16cm, trim=0 200 0 100]{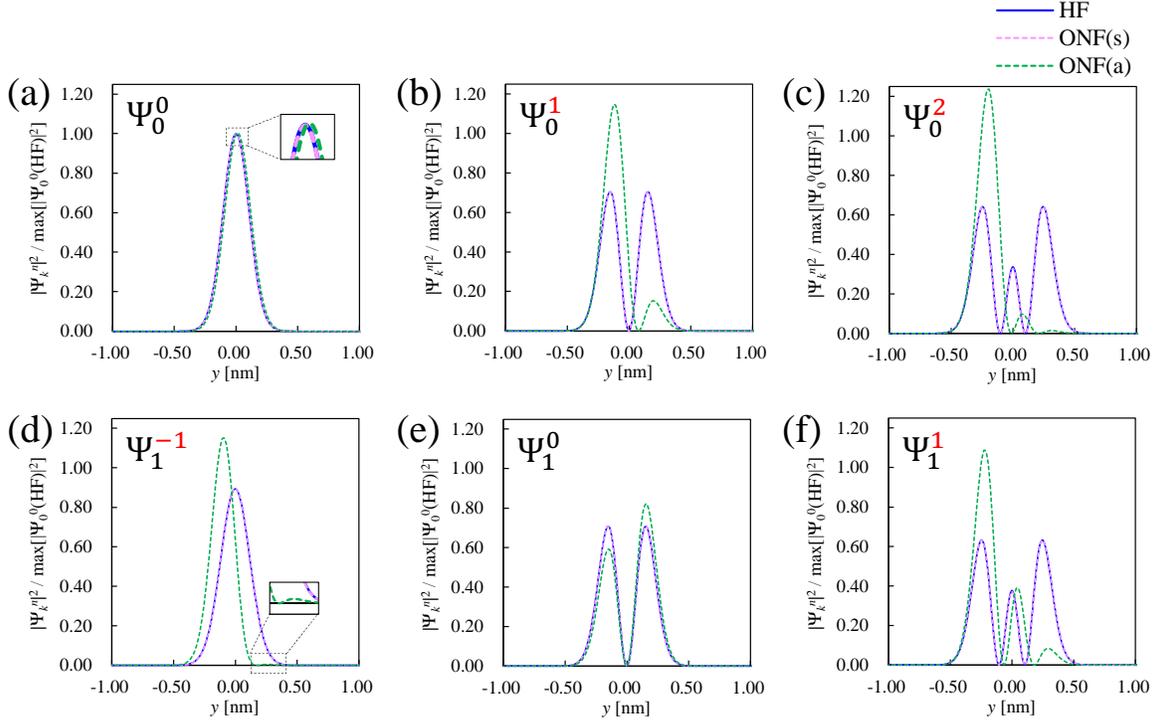}
\caption{(Color online) Probability density distributions of the dressed states $ \Psi_{k}^{n} $ for different values of $k$ and $n$ [($k$ = 0; $0 \le n \le 2$) and ($k$ = 1; $-1 \le n \le 1$)]. The probability density has been normalized by the highest value in $\Psi_{0}^{0}$ obtained by HF. The results obtained by HF, ONF(s), and ONF(a) (see text) are shown by the solid blue (black), broken pink (light gray), and broken green (dark gray) lines, respectively.
}\label{fig:dre_dis}
\end{figure}

The probability-density distributions of the dressed states $ \Psi_{k}^{n} $, that have contributed significantly to the spectra of Fig.~\ref{fig:power_spe_phi}, have been displayed in Fig.~\ref{fig:dre_dis} for $ k = 0 $  with $ n = 0\sim2 $ and $ k = 1 $ with $ n = -1\sim1 $. As for Figs.~\ref{fig:power_spe} and~\ref{fig:power_spe_phi}, each line corresponds to the result by HF, ONF(s), and ONF(a), respectively. The amplitude of $ \Psi_{k}^{n} $ here is normalized by $ \Psi_{0}^{0} $ obtained by HF. Although the eigenenergy $ \epsilon_{k}^{n} $ does hardly depends on the different excitation schemes, as has been mentioned, the wave function $ \Psi_{k}^{n} $ does, that is significantly affected by the spatial dependence of the fields defined by $ V_{s} $. 

The probability density distributions displayed in Figs.~\ref{fig:dre_dis}(a)~-~(c) represent the dressed states $ \Psi_{0}^{0\sim2} $ that originate from the ground state $ \psi_{0} $. The results for the one-photon and two-photon coupled state, $\Psi_{0}^{1}$ and $\Psi_{0}^{2}$, respectively, are significantly different from the zero-photon coupled state $\Psi_{0}^{0}$: The distributions for $\Psi_{0}^{1}$ and $\Psi_{0}^{2}$ have one and two nodes, respectively, indicating that they are influenced by the first- and second-excited state, $\psi_1$ and $\psi_2$, respectively, of the original electronic states owing to the near-resonance excitations. Further, the results by HF and ONF(s) are almost the same with each other while they show appreciable differences from those by ONF(a). 
In particular, in the cases of $ \Psi_{0}^{1, 2} $ (Figs. \ref{fig:dre_dis} (b) and (c)) the distributions by ONF(a) is asymmetric and is slanted to the left while those by HF and ONF(s) are symmetric with respect to the origin.

A similar observation can be made to the probability density distributions for $\Psi_{1}^{-1{\sim}1}$ shown in the lower panels of Figs.~\ref{fig:dre_dis}: First, the distributions of (d) and (f), representing the results for $\Psi_{1}^{-1}$ and $\Psi_{1}^{1}$, respectively, show different nodal structures than that of the zero-photon coupled states $\Psi_{1}^{0}$, that are apparently influenced significantly by the ground and second-excited states, $\psi_0$ and $\psi_2$, respectively, owing to the near-resonance excitations. Second, the distributions obtained by HF and ONF(s) are almost the same and symmetric with respect to the origin, while those obtained by ONF(s) are distinct than others and asymmetric. Another interesting observation can be made in Fig.~\ref{fig:dre_dis}(d) for $\Psi_{1}^{-1}$ where the distribution for ONF(a) has a node with an additional small peak at around $ y $ = 0.25 nm while those for HF and ONF(s) are singly-peaked without any node. These asymmetric distributions obtained by ONF(a) can be rationalized by that ONF(a) placed at 0.5 nm from the $ y $ axis produces an asymmetric local potential, that has deformed the wave function significantly. In other words the asymmetry of the ONF source has been inherited to the dressed states. Such inheritance of the asymmetry changes the parity of the probability density distribution of the dressed states, resulting in the change in their transition dipole moments. Consequently, the optical properties of the confined electron system under ONF(a) has been drastically modified.

\begin{figure}[htbp]
\centering
\includegraphics[width=16cm, trim=0 200 0 100]{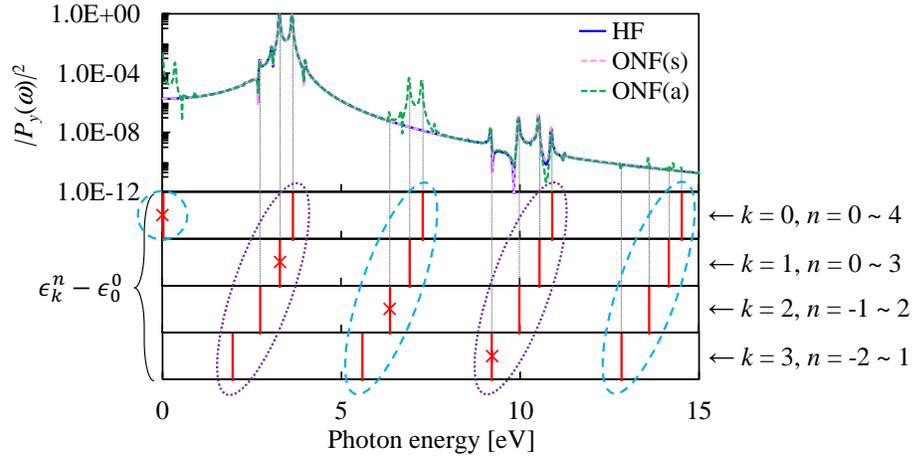}
\caption{(Color online) Normalized power spectrum of the induced dipole moment $ |P_{y}(\omega)|^{2} $ (the upper panel) and dressed energy $\epsilon_{k}^{n}$ measured from the energy of the reference state $\epsilon_{0}^{0}$, i.e., $\epsilon_{k}^{n} - \epsilon_{0}^{0}$ (the lower panels). For HF and ONF(s) excitations, the energy levels grouped by the purple (dark gray) dotted circles represent dressed states with odd parity, while those by the blue (light grey) circles represent states with even parity. See the captions to Figs. \ref{fig:power_spe} and \ref{fig:power_spe_phi} for further details.}\label{fig:power_spe_dre}
\end{figure}

Correspondence between the peaks in the spectra of the induced dipole moment $P_{y}$ displayed in Fig.~\ref{fig:power_spe} and the eigenenergies of the dressed states has been made in Fig.~\ref{fig:power_spe_dre}. The eigenenergies displayed in this figure have been subtracted by the energy of the reference state $\epsilon_{0}^{0}$, the zero-photon coupled ground state. Therefore, each peak in the spectra and the corresponding dressed state represent an optical transition from the electronic ground $\psi_0$ state. 

\begin{figure}[htbp]
\centering
\includegraphics[width=16cm, trim=0 250 0 100]{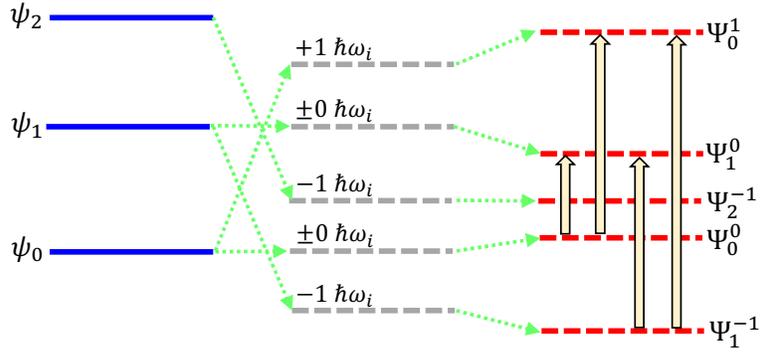}
\caption{(Color online) A schematic illustration of the formation of dressed states, $\Psi_{1}^{-1}$, $\Psi_{0}^{0}$, $\Psi_{2}^{-1}$, $\Psi_{1}^{0}$, and $\Psi_{0}^{1}$, from the original electronic states $\psi_0$, $\psi_1$, and $\psi_2$. A set of one-photon transitions involving the $\Psi_{0}^{0}$ state are indicated by arrows on the right-hand side of the figure.
}\label{fig:energylevels}
\end{figure}

For example, there are five peaks at around 3.6 eV with the highest double peaks assigned as $\Psi_{1}^{0}$ and $\Psi_{0}^{1}$, respectively. Four peaks out of them represent an one-photon excitation from $\psi_0$ to $\psi_1$, while the other one is a sequential two-photon excitation from $\psi_0$ to $\psi_2$ via $\psi_1$ in the weak-field limit. This situation is schematically illustrated in Fig.~\ref{fig:energylevels}: The energy levels of the original electronic states, $\psi_0$, $\psi_1$, and $\psi_2$ are displayed on the left-hand side of the figure. These energy levels generate zeroth-order dressed states, $\Psi_{1}^{-1}$, $\Psi_{0}^{0}$, $\Psi_{2}^{-1}$, $\Psi_{1}^{0}$, and $\Psi_{0}^{1}$, displayed in the middle of the figure, by adding or subtracting multiples of the photon energy $n{\hbar}\omega$. When the interaction $V_{s}$ is introduced, these zeroth-order dressed energy levels of the near-lying pairs of states, $(\Psi_{1}^{-1},\Psi_{0}^{0})$ and $(\Psi_{1}^{0},\Psi_{0}^{1})$ repel each other, making their energy gap larger as displayed on the right-hand side of the figure (Note that the level repulsion is illustrated in this figure rather exaggeratedly so as to explain the effect of the $V_s$ interaction. Indeed, with the current light intensity of ${\sim}10^{12}$ W/cm$^2$ the energies of the dressed eigenstates are almost the same as those of the corresponding zeroth-order states). Therefore, these four levels produce a total of four transition lines in the spectra with different energies as indicated by arrows in Fig.~\ref{fig:energylevels}. There is also another energy level corresponding to $\Psi_{2}^{-1}$, that lies closely. Since the second-excited $\psi_2$ state is dipole-allowed from $\psi_1$ but not from $\psi_0$, the optical transition from the ground $\psi_0$ state to this $\psi_2$ state occurs by a sequential two-photon excitation via $\psi_1$. This rationalizes the small amplitude of the spectral line corresponding to $\Psi_{2}^{-1}$ as can be seen in Fig.~\ref{fig:power_spe_dre}.

In Fig.~\ref{fig:power_spe_dre} a set of dressed states lying closely together and sharing the same parity within the framework of the symmetry-preserved excitation schemes of HF and ONF(s) have been grouped and indicated by dotted circles: the purple (dark gray) dotted circles indicate dressed states having odd parity while the watery dotted circles (light grey) indicate those having even parity. Since the $\Psi_{0}^{0}$ state is of parity even, this state can be optically coupled with states of odd parity by the HF and ONF(s) excitations, that are located in the spectra at around 3.6 eV and 11 eV corresponding, respectively, the first- and third-harmonic generations. Meanwhile, even parity states located at around 7.2, and 14 eV, corresponding, respectively, to SHG and FHG, would not be accessible by the HF and ONF(s) excitations. These optically-allowed and forbidden transitions can be known by calculating the transition dipole moment $ d $ defined by
\begin{align}
d_{kn,k'n'}=\int \Psi_{k}^{n} qy \Psi_{k'}^{n'} dv.
\label{tdp}
\end{align}
This definition makes it easy for us to know if the transition is allowed or not by considering simply their parity of the involved states in the case of the HF and ONF(s) excitations. In contrast, the ONF(a) excitation causes transitions from $ \Psi_{0}^{0} $ to almost all dressed states including transitions corresponding to SHG and FHG. As has been represented in Fig. \ref{fig:dre_dis}, the dressed states excited by ONF(a) have broken symmetry which give nonzero values of the transition dipole moment for much more combinations of states than do the cases by HF and ONF(s). Furthermore, although we focused on the transitions from $ \Psi_{0}^{0} $ in the above analysis, other dressed states also could be the origin of the spectra. In particular, as shown in Fig. \ref{fig:power_spe_phi}, $ \Psi_{1}^{-1\sim1} $ have made significant contributions in the time-dependent wave function. These states have given rise to, in addition to SHG and FHG, the DFG spectrum located at around 0.3 eV in Fig.~\ref{fig:power_spe_dre}. This is because the transition from $ \Psi_{0}^{n} $ to $ \Psi_{1}^{n-1} $ always involves the following energy difference:
\begin{align}
\epsilon_{0}^{n} - \epsilon_{1}^{n-1}=(\epsilon_{0}^{0}+n\hbar\omega_{i})-\{\epsilon_{1}^{0}+(n-1)\hbar\omega_{i}\}
=\hbar\omega_{i}-(\epsilon_{1}^{0}-\epsilon_{0}^{0})\approx 0.1\hbar\omega_{01}\approx 0.3 {\rm \, eV},
\label{dfg}
\end{align}
where we used the zeroth-order relation $ \epsilon_{1}^{0}-\epsilon_{0}^{0}\approx \hbar\omega_{01}$  and the factor 1.1 between $\omega_{01}$ and $\omega_i$ introduced in Eq. \eqref{td_vt}. This DFG is not allowed for the HF and ONF(s) excitations due to the parity restriction, but allowed for the ONF(a) excitation. 

%%%%%%%%%%%%%%%%%%%%%%%%%%%%%%%%%%%%%%%%%%%%%%%%%%%%%%%%%%%%%%%%%%%
\section*{IV. SUMMARY}
%%%%%%%%%%%%%%%%%%%%%%%%%%%%%%%%%%%%%%%%%%%%%%%%%%%%%%%%%%%%%%%%%%%

In the present study we have investigated the excitation dynamics of a finite electron system with an intense optical near-field (ONF) from a perspective of the dressed states. The system has been modeled by a single electron confined in a quasi-one-dimensional potential well, that mimics a quantum wire, while the ONF has been described by a short dipole source. The temporal evolution of the system has been calculated by solving the time-dependent Schr\"{o}dinger equation directly relying on the finite-difference time-domain method. We have examined three different light fields for exciting the system, namely, a spatially homogeneous field (HF) corresponding to a conventional plane-wave laser field, an inhomogeneous but symmetric ONF (ONF(s)), and an inhomogeneous and asymmetric ONF (ONF(a)). Optical responses of the system to these incident fields have been obtained by calculating its induced dipole moment from the time-dependent wave function. Fourier transformation of the resultant induced dipole moment has shown spectra of the radiation generated from the electron system through nonlinear interaction with the incident light fields. The spectra for HF and ONF(s) have indicated the first- and third-harmonic generations while the spectrum for ONF(a) has indicated not only these standard harmonic generations but also the second- and fourth-generations (SHG and FHG), and further, difference frequency generation (DFG).

To rationalize the resultant spectra, particularly, the appearance of the `forbidden' spectral lines excited by the inhomogeneous optical near field, we have constructed a theoretical model of dressed states that is applicable to systems subjected to an inhomogeneous ONF. The quasi-eigenstates obtained by solving the the Floquet matrix have shown that the dressed states under HF and ONF(s) are almost the same with each other in which they rigorously keep the parity of the original uncoupled electronic states. On other hand, in the case of ONF(a) the dressed states have shown probability density distributions that are asymmetric and thus breaking the symmetry of the original electronic states. This asymmetry of the probability density distributions of the dressed states inherited from ONF(a) has activated a number of otherwise forbidden transitions and thus allowed the new optical responses of SHG, FHG, and DFG. The present results have suggested that by creating and applying an asymmetric optical near-field to the target electron system we can access to `dark' states that are unreachable by conventional homogeneous light fields, enriching significantly information obtained by spectroscopic measurements. Results of such experiments can be predicted and thus designed by the present theoretical model of dressed states under an inhomogeneous optical near field.

\begin{acknowledgments}
This research was supported by JST-CREST under grant number JP-MJCR16N5. Theoretical computations were performed in part at the Research Center for Computational Science, Okazaki, Japan and Oakforest-PACS at the CCS, University of Tsukuba. T.S. acknowledges supports by Grants-in-Aid for Scientific Research (No.17H02725 and No.18K19065) from JSPS and by Nihon University CST Research Grants (2018).
\end{acknowledgments}

\bibliographystyle{apsrev4-1}
\bibliography{dressed.bib}
\newpage

%Fig.1%%%%%%%%%%%%%%%%%%%%%%%%%%%%%%%%%%%%%%%%%%%%%%%%%%%%%%%%%%%%%%%%%%

%Fig.2%%%%%%%%%%%%%%%%%%%%%%%%%%%%%%%%%%%%%%%%%%%%%%%%%%%%%%%%%%%%%%%%%%

%Fig.3%%%%%%%%%%%%%%%%%%%%%%%%%%%%%%%%%%%%%%%%%%%%%%%%%%%%%%%%%%%%%%%%%%

%Fig.4%%%%%%%%%%%%%%%%%%%%%%%%%%%%%%%%%%%%%%%%%%%%%%%%%%%%%%%%%%%%%%%%%%

%Fig.5%%%%%%%%%%%%%%%%%%%%%%%%%%%%%%%%%%%%%%%%%%%%%%%%%%%%%%%%%%%%%%%%%%

%Fig.6%%%%%%%%%%%%%%%%%%%%%%%%%%%%%%%%%%%%%%%%%%%%%%%%%%%%%%%%%%%%%%%%%%

%Fig.7%%%%%%%%%%%%%%%%%%%%%%%%%%%%%%%%%%%%%%%%%%%%%%%%%%%%%%%%%%%%%%%%%%

\end{document}